**Cross-classified multilevel models**

George Leckie

Centre for Multilevel Modelling and School of Education, University of Bristol

**Address for correspondence**

Centre for Multilevel Modelling

School of Education

University of Bristol

35 Berkeley Square

Bristol

BS8 1JA

United Kingdom

g.leckie@bristol.ac.uk

**Cross-classified multilevel models**

**Abstract**

Cross-classified multilevel modelling is an extension of standard multilevel modelling for non-hierarchical data that have cross-classified structures. Traditional multilevel models involve hierarchical data structures whereby lower level units such as students are nested within higher level units such as schools and where these higher level units may in turn be nested within further groupings or clusters such as school districts, regions, and countries. With hierarchical data structures, there is an exact nesting of each lower level unit in one and only one higher level unit. For example, each student attends one school, each school is located within one school district, and so on. However, social reality is more complicated than this, and so social and behavioural data often do not follow pure or strict hierarchies. Two types of non-hierarchical data structures which often appear in practice are cross-classified and multiple membership structures. In this article, we describe cross-classified data structures and cross-classified hierarchical linear modelling which can be used to analyse them.

**Introduction**

In cross-classified data there is not an exact nesting of each lower level unit in one and only one higher level unit. Rather, lower level units belong to pairs or combinations of higher level units formed by crossing two or more higher level classifications with one another. An example in educational research arises in studies of student attainment where students are nested within schools but are also nested within neighbourhoods. However, schools and neighbourhoods are not typically nested within one another as not all students from the same school live in the same neighbourhood nor do all students from the same neighbourhood attend the same school. Rather, schools and neighbourhoods are crossed with one another, with each student potentially belonging to any combination of school and neighbourhood. Students are described as nested within the cells of the two-way cross-classification of schools by neighbourhoods. An example in health services research arises in studies of hospital patient outcomes. Hospitals and general practitioners (GPs, i.e., family doctors) are cross-classified as GPs tend to refer their patients to different hospitals depending on patient need while hospitals typically treat patients who have been referred by many different GPs.

**The consequences of ignoring cross-classified structures**

It is important to incorporate cross-classified structures into our models when they arise in the data and are thought to lead to substantial clustering in the outcome under study. Ignoring cross-classified structures, by accounting for some nesting factors but not others, will typically lead us to overstate the relative importance of the factors that we do account for. This in turn may lead us to draw misleading conclusions about the relative importance of different sources of influence on the outcome (Luo and Kwok, 2009, 2012; Meyers and Beretvas, 2006). Thus, in our educational research example, accounting for schools, but ignoring neighbourhoods, will likely lead us to overestimate the importance of schools.

Similarly, in our health services research example, accounting for hospitals but ignoring GPs will likely lead us to overestimate the importance of hospitals.

**Model equations**

The cross-classified model for the above educational research example, where we adjust for a single covariate, can be written using "classification notation" (Browne et al. 2001) as

$$y_i = \beta_0 + \beta_1 x_i + u^{(2)}_{\text{school}(i)} + u^{(3)}_{\text{neigh}(i)} + e_i$$

$$u^{(2)}_{\text{school}(i)} \sim N(0, \sigma^2_{u(3)})$$

$$u^{(3)}_{\text{neigh}(i)} \sim N(0, \sigma^2_{u(2)})$$

$$e_i \sim N(0, \sigma^2_e)$$

where $y_i$ denotes the attainment of student $i$, $\beta_0$ is the model intercept, $x_i$ denotes the value of the covariate for that student, $\beta_1$ is the associated slope coefficient, $u^{(2)}_{\text{school}(i)}$ and $u^{(3)}_{\text{neigh}(i)}$ denote the school and neighbourhood random effects for that student, and $e_i$ denotes the student-level residual error. The subscripts $\text{school}(i)$ and $\text{neigh}(i)$ are 'classification functions' which return the school attended and neighbourhood resided in by student $i$, respectively. The (2) and (3) superscripts and subscripts are used to distinguish the different classifications from one another; convention has it that (1) superscripts and subscripts are not presented for $e_i$ and $\sigma^2_e$, but are implicit. The random effects and residual errors are assumed normally distributed with zero means and constant variances where $\sigma^2_{u(3)}$ denotes the between-school variance, $\sigma^2_{u(2)}$ denotes the between-neighbourhood variance, and $\sigma^2_e$ denotes the student-level residual error variance. The magnitudes of the variance components may then be compared to make statements about the relative contribution of each classification to the variation in the response, having adjusted for the covariate.

**Estimation and software**

Cross-classified models can be estimated by both frequentist (e.g., maximum likelihood) and Bayesian (e.g., Markov chain Monte Carlo, MCMC) estimation. Several software packages provide specific routines for fitting these models including the general-purpose packages R, SAS, SPSS, and Stata and the specialized multilevel modelling packages HLM and MLwiN (Charlton et al., 2019). MLwiN can be run from within both the R and Stata software (Leckie and Charlton, 2013; Zhang et al., 2016). For complex models, for example, with discrete responses or many different crossed classifications, Bayesian estimation will often be considerably more computationally efficient than frequentist estimation. Of the aforementioned packages, only MLwiN allows cross-classified models to be easily fitted by Bayesian methods (Browne, 2019, Chapter 15) in addition to frequentist methods (Rasbash et al., 2019, Chapter 18).

**Modelling extensions**

An important, but often overlooked, extension to cross-classified models is to allow for random interaction effects between the units of the different higher level classifications. This extension relaxes the assumption that the higher level units have additive effects. Note, however, that random interaction effects are not identified when there is only one observation per cell of the cross-classification as in this situation the random interaction effects will be confounded with the error. In our educational research example, school-neighbourhood combinations will often contain multiple students and so random interaction effects can be included. Doing so allows the effect that a student's school has on them to depend on the neighbourhood they live in and the effect that their neighbourhood has on them to depend on which school they attend. Including random interaction effects therefore allows for the fact that schools are likely to have different effects for students from different neighbourhoods and vice versa. When school and neighbourhood level variables are included in the model, we can choose to additionally include interactions between these variables to attempt to explain the variation in the random interaction effects across the cells of the cross-classification. Failure to account for random interaction effects will lead us to biased estimates of the other variance parameters included in the model (Shi et al., 2010).

The social and behavioural data that arise from social reality will often have far more complex data structures than those given in the educational and health services research examples above. To realistically model this complexity, we must often include further classifications in our models leading to three-way and four-way cross-classified models. In our educational research example, we may choose also to include the effects of schools from an earlier phase of schooling to account for potential carry-over effects of these schools on student attainment. Students are then nested within the cells of a three-way cross-classification of high schools by junior schools by neighbourhoods. Further hierarchical structures may also need to be incorporated into the model. For example, if our educational data are from an international comparative study, we may want to incorporate country effects on student attainment. Very few students will move countries and so the high schools, junior schools, and neighbourhoods are nested within countries. In our health services research example, we could include ward effects as a third cross-classifying factor to account for within-hospital-between-ward variation. Were we to have repeated measurements on patients' outcomes during their stay in hospital, we would add measurement occasion as a new lowest level to the model.

The last two examples have shown how complex multilevel models can become when we try to extend them to realistically reflect the complex data structures that arise in social reality. Unit diagrams and classification diagrams have both been proposed as helpful aides to understanding and communicating complex multilevel data structures (Browne et al., 2001). Similarly, classification notation, which avoids the proliferation of subscripts that arises when we combine many different data structures in a single model, has been proposed as an alternative to standard notation when expressing these models in equation form (Browne et al., 2001).

An interesting use of cross-classified models is in panel data. In multilevel analysis, most panel data is treated as two-level where time is nested within panels. This is the case in individual panel surveys where measurement occasions are nested within individuals. However, in longer panels where there are many time points and where we might expect the outcome to vary systematically from time point to time point, we may treat the panels as cross-classified with time. An example is a state-year unemployment panel where we could chose to treat the unemployment measurements as nested within the cells of a cross-classification of states by years. If the panel is balanced, there will be exactly one observation in every cell of the classification. If the panel is unbalanced panel, for example, because unemployment counts were not returned by certain states in a particular year, the associated cells of the cross-classification will be empty. Note that panel data is an example of a cross-classified data structure where it is not possible to identify random interaction effects between the cross-classifying units as there is a maximum of one observation per cell. One could potentially resolve this problem by collecting county level unemployment data within each state-year combination to ensure that we have multiple measurements per cell. One concern with including state and county effects in this example is that each set of effects may be spatially correlated. One solution is to explicitly model these dependencies by including a multiple membership structure in the cross-classification.

Cross-classified models can also be extended to handle dyadic data (Kenny and Kashy 2011), for example family data on the relationship quality between each pair of family members, or migration and trade flow data between geographic areas. So, for migration flows between areas, the response observations are the measured flows and these are modelled as being simultaneously nested within areas in their roles as origins and areas in their role as destinations. Thus, every area has both an origin and destination random effect and these effects are allowed to correlate. Every pair of areas then has a pair of correlated residuals errors measuring how the observed flows deviate from those predicted solely by the origin and destination covariates and the area origin and destination random effects.

**Further reading**

Introductory, intermediate, and advanced treatments of multilevel models are given in the multilevel modelling textbooks by Snijders and Bosker (2012), Raudenbush and Bryk (2002), and Goldstein (2011), respectively. Accessible introductions to cross-classified models are given by the report by Fielding and Goldstein (2006) and the book chapter by Beretvas (2010). More advanced treatments of cross-classified models are provided in the multilevel textbook by Goldstein (2011, Chapter 12), the book chapters on cross-classified models by Rasbash and Browne (2001, 2008), the paper by Browne et al. (2001), and the reports by Fielding et al. (2006) and Leckie (2013). Examples of cross-classified hierarchical linear modelling in applied research can be seen in Leckie (2009), Rasbash et al, (2010), and Raudenbush (1993) who all model student attainment accounting for the nesting of students within the cross-classification of schools by neighbourhoods. Rasbash and Goldstein (1994) model student attainment accounting for the nesting of students within the cross-classification of primary by secondary schools. Leckie and Baird (2011) and Baird, Meadows, Leckie and Caro (2017) model the scores awarded by raters to students' essays. The scores are cross-classified by raters and students as each rater scores every student and each student is scored by every rater. Moving beyond education, Brunton-Smith, Sturgis and Leckie (2017, 2018) present two applications to survey respondents' attitudes where they model respondents as nested within a cross-classification of neighbourhoods and interviewers. In terms of using extended cross-classified models to fit dyadic data, Browne, Leckie, Prime, Perlman and Jenkins (2016) present an application to sensitivity shown between family members, Koster and Leckie (2016) present an application to altruism between villagers, and Zhang et al. (2018) present an application to migration flows between Chinese provinces.

CROSS-CLASSIFIED MULTILEVEL MODELSBrunton-Smith, I., Sturgis, P., Leckie, G. (2017). Detecting, and understanding interviewer effects on survey data using a cross-classified mixed-effects location scale model. Journal of the Royal Statistical Society: Series A (Statistics in Society), 180, 2, 551-568.

Brunton-Smith, I., Sturgis, P., & Leckie, G. How collective is collective efficacy? The importance of consensus in judgments about community cohesion and willingness to. Criminology, 56, 608-637.

Charlton, C., Rasbash, J., Browne, W.J., Healy, M. and Cameron, B. (2019) *MLwiN Version 3.04*. Centre for Multilevel Modelling, University of Bristol.

Fielding, A., & Goldstein, H. (2006). *Cross-classified and multiple membership structures in multilevel models: An introduction and review*. Research Report RR791. London: Department for Education and Skills. Retrieved April 4, 2013 from http://www/bristol.ac.uk/cmm/team/cross-classified-review.pdf

Fielding, A., Thomas, H., Steele, F., Browne, W., Leyland, A., Spencer, N., et al. (2006). *Using cross-classified multilevel models to improve estimates of the determination of pupil attainment: A scoping study*. Research report. Birmingham: School of Education, University of Birmingham. Retrieved April 4, 2013 from http://www.bristol.ac.uk/cmm/team/dfes-scoping-report.pdf

Goldstein, H. (2011). *Multilevel statistical models* (4th ed.). London: Wiley.

Kenny, D. A. and Kashy, D. A. (2011) 'Dyadic Data Analysis Using Multilevel Modeling', in Handbook of Advanced Multilevel Analysis, pp. 344–60.

Koster, J., & Leckie, G. (2014). Food sharing networks in lowland Nicaragua: An application of the social relations model to count data. Social Networks, 38, 100-110.
10